\documentclass[12pt]{article}
\usepackage{amssymb}
\usepackage{epsfig}
\begin{document}
%
\setlength{\baselineskip}{0.65cm}
\setlength{\parskip}{0.35cm}
\renewcommand{\thesection}{\Roman{section}}
%
\begin{titlepage}
%
\begin{flushright}
BNL-NT-03/20 \\
RBRC-333 \\
September 2003
\end{flushright}

\vspace*{1.1cm}
\begin{center}
\LARGE

{\bf {Longitudinally Polarized Photoproduction of}}\\

\medskip
{\bf {Inclusive Hadrons Beyond the Leading Order}}\\

\vspace*{2.5cm}
\large 
{B.\ J\"{a}ger$^{a}$, M.\ Stratmann$^{a}$, and W.\ Vogelsang$^{b,c}$}

\vspace*{1.0cm}
\normalsize
{\em $^a$Institut f{\"u}r Theoretische Physik, Universit{\"a}t Regensburg,\\
D-93040 Regensburg, Germany}\\

\vspace*{0.5cm}
{\em $^b$Physics Department, Brookhaven National Laboratory,\\
Upton, New York 11973, U.S.A.}\\

\vspace*{0.5cm}
{\em $^c$RIKEN-BNL Research Center, Bldg. 510a, Brookhaven 
National Laboratory, \\
Upton, New York 11973 -- 5000, U.S.A.}
\end{center}

\vspace*{2.0cm}
\begin{abstract}

\noindent
We present a complete next-to-leading order QCD calculation for 
single-inclusive large-$p_T$ hadron production in 
longitudinally polarized lepton-nucleon collisions, 
consistently including ``direct'' and ``resolved'' photon contributions.
This process could be studied experimentally at a future
polarized lepton-proton collider like eRHIC at BNL.
We examine the sensitivity of such measurements
to the so far completely unknown parton content
of circularly polarized photons.
\end{abstract}
\end{titlepage}
\newpage
%
\section{Introduction}
%
The successful start of the RHIC spin program marks a new era in spin 
physics. Very inelastic $pp$ collisions at high energies open up unequaled 
possibilities to thoroughly address many interesting questions, most 
importantly perhaps those concerning the still unknown role of polarized 
gluons \cite{ref:rhic}. QCD analyses of upcoming high-$p_T$ prompt photon, 
jet, heavy flavor, inclusive-hadron, and $W^\pm$ boson production data 
will greatly deepen and enrich the understanding of the nucleon and of QCD
that we had gained previously from polarized deeply-inelastic 
scattering (DIS) of leptons off nucleons. 

Investigations of interactions between polarized leptons and 
nucleons will, however, continue to play a vital role in spin 
physics. Besides the many observables of DIS, 
{\em photoproduction} processes are particularly interesting. 
They give access to the parton content of circularly polarized 
photons, due to the presence of contributions to the cross 
section where the photon ``resolves'' into its parton content, for 
instance, by fluctuating into a vector meson of the same quantum 
numbers,  prior to the hard QCD interaction. Such contributions
compete with the ``direct'' part, for which the photon simply interacts 
as an elementary pointlike particle. 

It is well-known \cite{ref:owens} that at lower (fixed-target) energies, 
the direct component tends to dominate over the resolved one. 
Only at collider energies do resolved contributions become
important. Extensive studies in the unpolarized case
at the HERA and LEP colliders have firmly
established the concept of photonic parton densities \cite{ref:klasen}
and shown that resolved photon processes contribute a sizable fraction of 
photoproduction events in certain kinematic regions. The fact that
$e^+e^-$ and $ep$ data can be universally described by the same set 
of photonic parton densities marks an important test and success of QCD.

It has not yet been possible to observe resolved contributions for 
{\em polarized} photons, due to the fact that only fixed-target
lepton-nucleon experiments have been performed so far in the polarized
case. In this context, an exciting possibility would be to have a polarized
lepton-proton {\em collider} such as the planned eRHIC project at 
BNL \cite{ref:eic}, which is currently under discussion.

Motivated by this prospect, we present in this paper a study for
photoproduction of single-inclusive high-$p_T$ hadrons at a future 
lepton-proton collider with longitudinally polarized beams, and
investigate the sensitivity of this reaction to the contribution 
from resolved photons. We had made first exploratory studies of 
this, and related, reactions in \cite{ref:lostudies}. Our 
present study goes beyond the results of \cite{ref:lostudies} in
that we now consistently include both the direct and the resolved photon
processes in the next-to-leading order (NLO) of QCD. This is crucial for 
reliable quantitative analyses in the future as theoretical uncertainties 
are expected to be under much better control beyond the lowest order (LO) 
approximation of QCD. For the resolved contribution, the NLO 
corrections to polarized QCD hard-scattering have become available only 
recently in \cite{ref:nlopion,ref:danielpion}. For the polarized direct component, the 
corresponding cross sections were already calculated some time ago
in \cite{ref:nlopoldir}. 
We have rederived the expressions and agree with the published results.
In addition to extending our study to NLO, we also tailor our 
phenomenological results to the eRHIC situation. 

In the next section we sketch the technical framework for 
the description of polarized photoproduction reactions.
Section III is devoted to numerical studies for the eRHIC.
We conclude in Section IV.

\section{Technical Framework}
%
We consider the spin-dependent photoproduction cross section
for the  reaction $lp\to l^{\prime} H X$, where the hadron $H$ is at high
transverse momentum $p_T$, ensuring a large momentum transfer. 
We may then write the differential cross section in a factorized form:
\begin{eqnarray}
\frac{d\Delta\sigma}{dp_Td\eta} &=& 
\frac{1}{2}\left[\frac{d\sigma_{++}}{dp_Td\eta}-
\frac{d\sigma_{+-}}{dp_Td\eta}\right] 
\nonumber \\[3mm]
&=& \frac{2p_T}{S} \sum_{a,b,c} \int^1_{1-V+VW} 
\frac{dz}{z^2} \int_{VW/z}^{1-(1-V)/z}\!\!\!\frac{dv}{v(1-v)} 
\int^1_{VW/vz}\!\!\frac{dw}{w}\Delta f^l_a(x_l,\mu_f) \Delta f^p_b(x_p,\mu_f) 
D_c^{H}(z,\mu_f')  
\nonumber \\[3mm]
&\times&  \,\left[ 
\frac{d\Delta \hat{\sigma}^{(0)}_{ab\rightarrow cX}(s,v)}{dv} 
\delta (1-w) + \frac{\alpha_s(\mu_r)}{\pi} \, \frac{d\Delta 
\hat{\sigma}^{(1)}_{ab\rightarrow cX}(s,v,w,\mu_r,\mu_f,\mu_f')}{dvdw}
\right] \label{eq:eq2} \;\; . 
\end{eqnarray}
This generic form applies to both the direct and the resolved
cases. The subscripts ``$++$'' and ``$+-$'' 
denote the helicities of the colliding lepton and proton,
$\mu_r$ the renormalization scale, 
and $S\equiv (P_l+P_p)^2$ the available c.m.s.\ energy squared.
Neglecting all masses, $V$ and $W$ in Eq.~(\ref{eq:eq2}) can be 
expressed in terms of the c.m.s.\ pseudorapidity $\eta$ and transverse 
momentum $p_T$ of the observed hadron:
\begin{equation}
V=1-\frac{p_T}{\sqrt{S}} e^{\eta}\;\;\;\mathrm{and}\;\;\;
W=\frac{p_T^2}{S V (1-V)}\;\;\;.
\label{eq:eq3}
\end{equation}
We count positive rapidity in the proton's forward direction.

\begin{figure}[t]
\begin{center}
\begin{minipage}{7cm}
\epsfig{figure=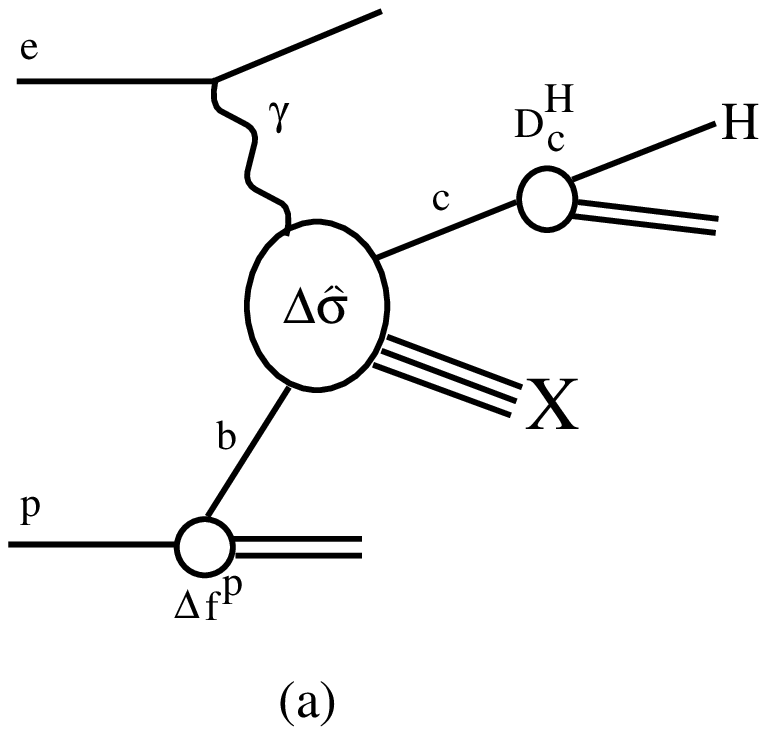,width=0.95\textwidth}
\end{minipage}
\begin{minipage}{7cm}
\epsfig{figure=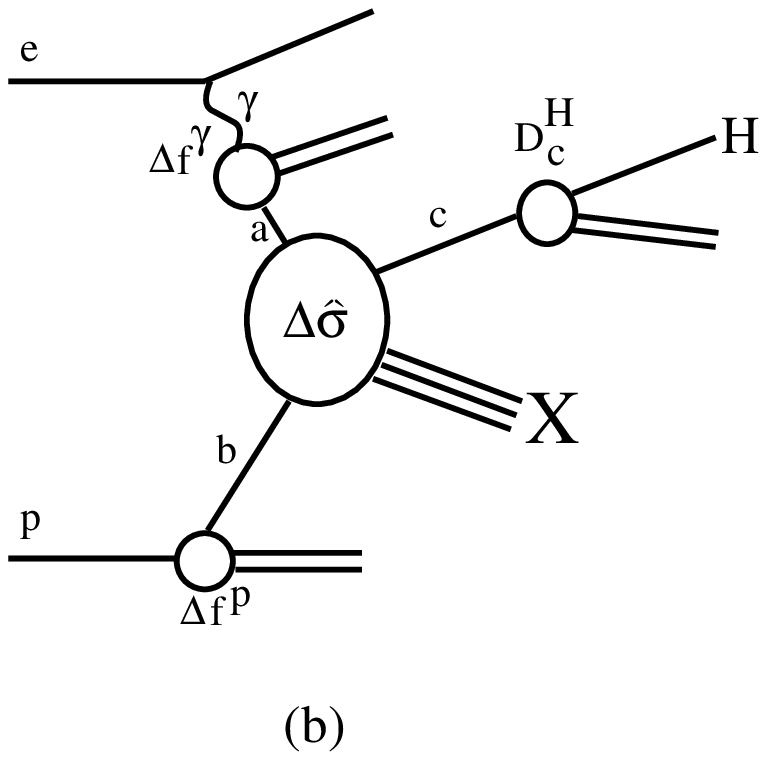,width=0.95\textwidth}
\end{minipage}
\end{center}
\vspace*{-0.5cm}
\caption{\sf Generic direct (a) and resolved (b) photon contributions to 
the process $lp\rightarrow l^{\prime}HX$. \label{fig:fig1}}
\end{figure}

The sum in Eq.~(\ref{eq:eq2}) is over all partonic channels $a+b\to c+X$
contributing to the single-inclusive cross sections, with 
$d\Delta \hat{\sigma}_{ab\rightarrow cX}^{(0)}$ and
$d\Delta \hat{\sigma}_{ab\rightarrow cX}^{(1)}$
the associated polarized LO and NLO partonic cross sections, 
respectively. For the direct contribution to the cross section,
depicted in Fig.~\ref{fig:fig1} (a), we have $a=\gamma$, while for 
the resolved ones [see Fig.~\ref{fig:fig1} (b)]
$a$ stands for the parton emerging from the photon. 
For convenience, we introduce effective polarized parton 
densities, $\Delta f_a^l(x_l,\mu_f)$, in the lepton, with $x_l$ the fraction
of the lepton momentum carried by parton $a$ and $\mu_f$ the 
factorization scale. The $\Delta f_a^l(y,\mu_f)$ are defined via 
the convolution
\begin{equation}
\label{eq:eq7}
\Delta f_a^l(x_l,\mu_f) = \int_{x_l}^1 \frac{dy}{y} \Delta P_{\gamma l} (y)
\Delta f_a^{\gamma}\left(x_{\gamma}=\frac{x_l}{y},\mu_f\right)\;\;,
\end{equation}
with 
\begin{equation}
\Delta P_{\gamma l} (y) =
\frac{\alpha_{em}}{2\pi} \left\{ \left[
\frac{1-(1-y)^2}{y}\right]
\ln \frac{Q_{\max}^2 (1-y)}{m_l^2 y^2} + 2 m_l^2 y^2 \left( 
\frac{1}{Q^2_{\max}}-\frac{1-y}{m_l^2 y^2} \right) \right\}
\label{eq:eq6}
\end{equation}
being the spin-dependent Weizs\"{a}cker-Williams ``equivalent photon'' 
spectrum including its non-logarithmic contributions\footnote{The latter
turn out to be numerically unimportant in our study, but we keep
them for completeness.} \cite{ref:daniel}. 
Here, $m_l$ is the lepton mass and 
$Q_{\max}$ the allowed upper limit on the radiated photon's virtuality, 
to be fixed by the experimental conditions. For the direct case, one
simply replaces
\begin{equation}
\Delta f_a^{\gamma}(x_{\gamma},\mu_f)\;\to\;\delta(1-x_{\gamma})\; ,
\label{eq:eq7a}
\end{equation}
while in the resolved contribution the yet unmeasured parton distributions 
of circularly polarized photons occur that are the focus of this paper. 
Likewise, $\Delta f_b^p(x_p,\mu_f)$ denotes the density of 
parton $b$ in the polarized proton, and $D_c^{H}(z,\mu_f')$
is the usual unpolarized fragmentation function
for parton $c$ going into the observed hadron $H$, at scale $\mu_f'$.
The relations between parton-level and hadron-level variables are
\begin{equation} \label{eq:eq5}
s\equiv (p_a+p_b)^2=x_l x_p S \;\; , \;\;
x_l = \frac{VW}{vwz} \;\; , \;\;\;\; x_p = \frac{1-V}{z (1-v)} \;\;.
\end{equation}
%

%

As we mentioned earlier, it is a main feature of our paper that
for the first time we include both the direct and the resolved 
contributions to the polarized cross section at NLO, as indicated by
the first-order corrections $d\Delta \hat{\sigma}_{ab\rightarrow cX}^{(1)}$
in Eq.~(\ref{eq:eq2}). The actual calculation of these corrections 
is generally a formidable task. Since later on we will mainly 
consider spin asymmetries, we also need the corresponding spin-averaged 
partonic cross sections. Let us go through the various ingredients 
that we need in a little more detail.

The resolved contribution is technically equivalent to a 
hadroproduction cross section such as $pp\to HX$. 
The corresponding $d\Delta \hat{\sigma}_{ab\rightarrow cX}^{(1)}$
may therefore be immediately adopted from our recent calculation 
\cite{ref:nlopion} for this case. In that reference, we also 
re-derived the (previously known \cite{ref:aversa}) expressions
for the unpolarized case. There are in total 16 different 
partonic channels. 

The NLO contributions to the direct-photon component have been
calculated in \cite{ref:aurenche,ref:nlounpdir} and \cite{ref:nlopoldir}
for the unpolarized and polarized cases, respectively. Having the
techniques available that we used in \cite{ref:nlopion}, it was a 
straightforward, though tedious, work to check the old results from
the scratch. For the polarized case we fully agree with \cite{ref:nlopoldir}, 
at an analytical level. In the unpolarized case, we were able to 
identify some minor mistakes (beyond those noted in \cite{ref:nlopoldir})
in the computer code of \cite{ref:nlounpdir}; apart from those we agree 
with that calculation as well. 

The technical details of the NLO calculation for the direct part 
may be found in \cite{ref:nlopoldir}. We will not repeat these here, 
but will only recall an important aspect of photoproduction cross
sections beyond the LO: writing the cross section as the sum
of the direct and the resolved photon contributions,
\begin{equation}
d\Delta \sigma = d\Delta \sigma_{\mathrm{dir}} + d\Delta 
\sigma_{\mathrm{res}}\;\;,
\label{eq:eq8}
\end{equation}
we emphasize that neither $d\Delta 
\sigma_{\mathrm{dir}}$ nor $d\Delta \sigma_{\mathrm{res}}$ are 
{\em individually} physical, i.e., experimentally measurable.
This can be easily seen by noticing that the $2\to 3$ contributions 
to the direct cross section have singular configurations arising 
from a collinear splitting of the incoming photon into a quark-antiquark 
pair. These contributions are absorbed, according to the factorization
theorem, into the so-called ``pointlike'' part of the NLO photonic 
parton densities and thus into the {\em resolved} component. 
The subtraction of the singular part is unique only 
up to finite pieces, and the ensuing ambiguity (i.e., the 
choice of ``factorization scheme'') cancels only when 
the direct and resolved parts are combined.  

As an example, when using dimensional regularization ($d=4-2\varepsilon$),
the appropriate ``counter cross section'' that cancels the
$\gamma\to q\bar{q}$ collinear singularity in the subprocess 
$\gamma \bar{q} \rightarrow gX$ is schematically given by
\begin{equation}
\frac{d\Delta \hat{\sigma}_{\gamma \bar{q}\rightarrow gX}^{\mathrm{counter}}}
{dvdw} \sim
-\frac{\alpha_{em}}{2\pi}\;  \Delta H_{q\gamma}\otimes
d\Delta \hat{\sigma}_{q \bar{q}\rightarrow gg}\;
\label{eq:eq9}
\end{equation}
where $d\Delta \hat{\sigma}_{q \bar{q}\rightarrow gg}$ is the polarized
Born cross section for the reaction $q\bar{q}\rightarrow gg$
in $d$ dimensions, and ($e_q$ denotes the quark charge)
\begin{equation}
\Delta H_{q\gamma}(x,\mu_f)= 
\left(-\frac{1}{\varepsilon} + \gamma_E - \ln 4\pi \right)\,
3 e_q^2 \left[2x-1\right]\,
\left( \frac{s}{\mu_f^2} \right)^\varepsilon + \Delta h_{q\gamma}(x)\;\;.
\label{eq:eq10}
\end{equation}
The freedom in the choice of the factorization prescription is
reflected by the arbitrary finite piece $\Delta h_{q\gamma} (x)$ 
which may be subtracted alongside the singular $1/\varepsilon$ 
contribution. For the  $\overline{\rm{MS}}$ convention, 
$\Delta h_{q\gamma} (x)=0$. An alternative factorization scheme 
($\mathrm{DIS}_{\gamma}$) was proposed in the unpolarized case 
\cite{ref:grvdisgamma} to facilitate the analysis of data on the 
DIS photon structure function $F_2^{\gamma}$. Here the photonic Wilson 
coefficient $C_{2,\gamma}$ in the NLO expression for $F_2^{\gamma}$ is 
absorbed into the definition of the quark distributions of the photon.
A similar scheme can be set up in the polarized case as 
well \cite{ref:polphotnlo} by subtracting the corresponding polarized 
coefficient function $\Delta C_{1,\gamma}$ in the structure function 
$g_1^{\gamma}$. The photonic parton densities in the $\overline{\mathrm{MS}}$ 
and $\mathrm{DIS}_{\gamma}$ schemes are then related in the
following way \cite{ref:polphotnlo}:
\begin{equation}
\Delta f_a^{\gamma,\overline{\mathrm{MS}}}(x,\mu_f) =
\Delta f_a^{\gamma,\mathrm{DIS}_{\gamma}}(x,\mu_f) + 
\delta \Delta f_a^{\gamma}(x)\;\;,
\label{eq:eq11}
\end{equation}
where
\begin{eqnarray}
\nonumber
\delta \Delta  f_q^{\gamma}(x)&=&
\delta \Delta  f_{\bar{q}}^{\gamma}(x)=-2 N_C e_q^2 \frac{\alpha_{em}}{4\pi}
\left[ (2x-1) \left( \ln \frac{1-x}{x} -1 \right) +2(1-x)\right]\; , \\
\delta \Delta f_g^{\gamma}(x)&=&0\;\;.
\label{eq:eq12}
\end{eqnarray}
The relation between the partonic cross sections for the direct 
contribution reads accordingly:
\begin{equation}
d\Delta \hat{\sigma}^{\mathrm{DIS}_{\gamma}}_{\gamma b \rightarrow c X} =
d\Delta \hat{\sigma}^{\overline{\mathrm{MS}}}_{\gamma b \rightarrow c X} +
\sum_a \delta \Delta f_a^{\gamma} \otimes 
d\Delta \hat{\sigma}_{a b \rightarrow c X} 
\label{eq:eq13}
\end{equation}
with $\delta \Delta f_a^{\gamma}$ given above and the 
symbol $\otimes$ denoting a standard convolution. The 
$d\Delta \hat{\sigma}_{a b \rightarrow c X}$
are the appropriate LO partonic cross sections.
For the resolved cross section all partonic cross sections  
remain unaffected by this transformation. It is 
straightforward to verify that the scheme transformation indeed
leaves the physical cross section in Eq.~(\ref{eq:eq8})
invariant.

Our explicit analytical expressions for the polarized NLO subprocess cross
sections for the direct and resolved cases are
too lengthy to be given here, but can be found in our computer 
code which is available upon request.

\section{Numerical Results}
%
Let us now turn to a first numerical application of our
analytical results. Instead of presenting a full-fledged phenomenological 
study of single-inclusive hadron production in polarized $lp$ collisions, 
we only focus on the most interesting questions: the importance of the NLO
corrections, the dependence on unphysical scales, and the sensitivity to
the yet unmeasured parton densities of circularly polarized photons.

For all our calculations we choose a c.m.s.\ energy of $\sqrt{S}=100\,
\mathrm{GeV}$ as planned for polarized electron-proton collisions at eRHIC. 
We limit ourselves to the case of neutral pions ($\pi^0$); extension
to other hadrons would be straightforward. We will always perform the 
NLO (LO) calculations using NLO (LO) parton distribution functions, 
fragmentation functions, and the two-loop (one-loop) expression for 
$\alpha_s$. The value for $\alpha_s$ is taken according to our choice of
proton parton distributions, which is CTEQ5 \cite{ref:cteq5}
for the unpolarized case and GRSV \cite{ref:grsv} for the
polarized one. 

In the equivalent photon spectrum in Eq.~(\ref{eq:eq6}) we use
similar parameters as the H1 and ZEUS experiments at HERA. We
employ $Q_{\max}^2=1\,\mathrm{GeV}^2$ and take the photon 
momentum fraction to be within $0.2\le y \le 0.85$. 

The fragmentation functions for neutral pions, $D^{\pi}_c$, have been 
determined with some accuracy from analyses of data for pion production in
$e^+e^-$ collisions \cite{ref:kkp,ref:kretzer}. Even though our knowledge 
of the $D_c^{\pi}$ certainly still needs improvement, we assume
for this study that the fragmentation functions will be well
known by the time experiments at a future polarized lepton-proton collider
are carried out. For our analysis we take the LO and NLO sets of 
Ref.~\cite{ref:kkp} which also yield good agreement with
recent high-$p_T$ data for $pp\to\pi^0 X$ at $\sqrt{S}=200\,
\mathrm{GeV}$ by the PHENIX collaboration \cite{ref:phenixpion}.

\begin{figure}[t]
\vspace*{-1.3cm}
\begin{center}
\epsfig{figure=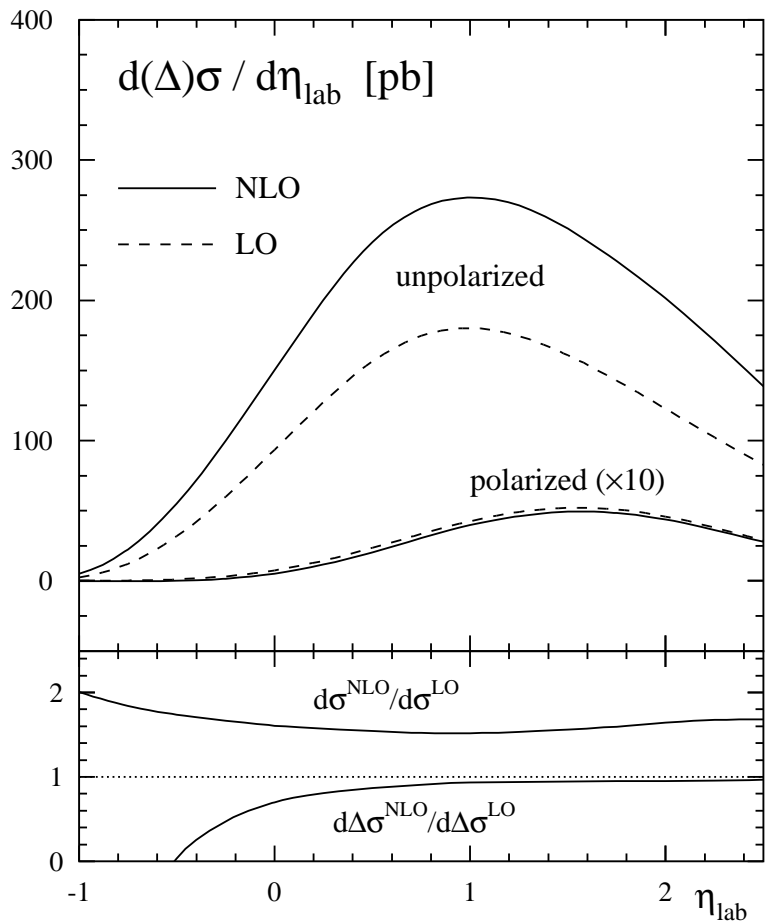,width=0.70\textwidth}
\end{center}
\vspace*{-1.0cm}
\caption{\sf Unpolarized and polarized $\pi^0$ photoproduction cross 
sections in NLO (solid lines) and LO (dashed lines) at $\sqrt{S}=100$ GeV. 
The polarized results are enlarged by a factor of 10. The lower panel 
shows the ratios of the NLO and LO results in each case. \label{fig:fig2}}
\end{figure}
Since the parton distributions $\Delta f^{\gamma}$ of circularly polarized 
photons are completely unmeasured so far, we have to invoke some model for 
them. Two extreme scenarios were considered in 
\cite{ref:lomodels,ref:lostudies}, with ``maximal'' 
[$\Delta f^{\gamma} (x,\mu_0) = f^{\gamma} (x,\mu_0)$] and ``minimal'' 
[$\Delta f^{\gamma} (x,\mu_0) = 0$] saturation of the ``positivity 
constraint'' $|\Delta f^{\gamma}(x,\mu_0)| \le f^{\gamma}(x,\mu_0)$. 
$\mu_0$ denotes the scale where the boundary conditions \cite{ref:lomodels} 
for the evolution are specified. For the spin-averaged densities 
$f^{\gamma}$ the phenomenologically successful GRV photon distributions 
\cite{ref:grvphot} were used. We employ these two extreme sets for the 
$\Delta f^{\gamma}$ in Eq.~(\ref{eq:eq7}), along with appropriate LO 
\cite{ref:lomodels} and NLO \cite{ref:polphotnlo,ref:sieg} boundary 
conditions and evolution equations. Our default choice for the 
numerical studies will be the ``maximal'' set.

As we already pointed out in previous LO studies \cite{ref:lostudies}, 
for single-inclusive measurements rapidity differential cross sections 
are particularly suited for extracting the resolved contributions 
and hence the polarized photonic parton distributions. 
This can be easily understood by looking at the 
momentum fractions of the photon and proton distributions as a function
of rapidity. As we count positive rapidity in the proton's forward direction,
large $x_{\gamma}\rightarrow 1$ are probed at large negative rapidities. 
Here one expects no difference between the two photon scenarios since
the cross section is dominated by the direct photon contribution
sitting strictly at $x_{\gamma}= 1$.
In addition, in the region $x_{\gamma} \rightarrow 1$ the 
$\Delta f^{\gamma}$ are dominated by the ``pointlike'' part which only 
depends on the starting scale $\mu_0$ but not on the details of the 
hadronic input. On the other hand, for large and positive rapidities 
small $x_{\gamma}$ values are probed, and the resolved contribution is 
expected to dominate more and more. 

Figure~\ref{fig:fig2} shows our results for the unpolarized and polarized
photoproduction cross sections for neutral pions at LO and NLO 
at $\sqrt{S}=100\,\mathrm{GeV}$ as functions of the
laboratory-frame pseudorapidity $\eta_{\mathrm{lab}}$. For the
asymmetric eRHIC kinematics ($E_p=250\,\mathrm{GeV}$,
$E_e=10\,\mathrm{GeV}$), $\eta_{\mathrm{lab}}$ is related 
to the c.m.s.\ frame rapidity via
\begin{equation}
\eta_{\mathrm{lab}} = \eta + \frac{1}{2} \ln \frac{E_p}{E_e}\;\;\;.
\end{equation}
We have integrated over the transverse momentum $p_T\geq 4\,
\mathrm{GeV}$ of the produced pion. All scales have been set
equal: $\mu_r=\mu_f=\mu_f'=p_T$.

The lower part of the figure displays the so-called ``$K$-factor''
\begin{equation}
K=\frac{d(\Delta)\sigma^{\rm NLO}}{d(\Delta)\sigma^{\rm LO}} \;\; .
\end{equation}
One can see that in the unpolarized case the corrections are roughly 
constant, of size $60\%-80\%$ over most of the $\eta_{\mathrm{lab}}$ 
region considered. In the polarized case, we find generally much smaller 
corrections, at least for the above choice of scales. 
Only for negative  $\eta_{\mathrm{lab}}$ where the cross 
section changes sign are the corrections large.

%
\begin{figure}[t]
\vspace*{-1.2cm}
\begin{center}
\epsfig{figure=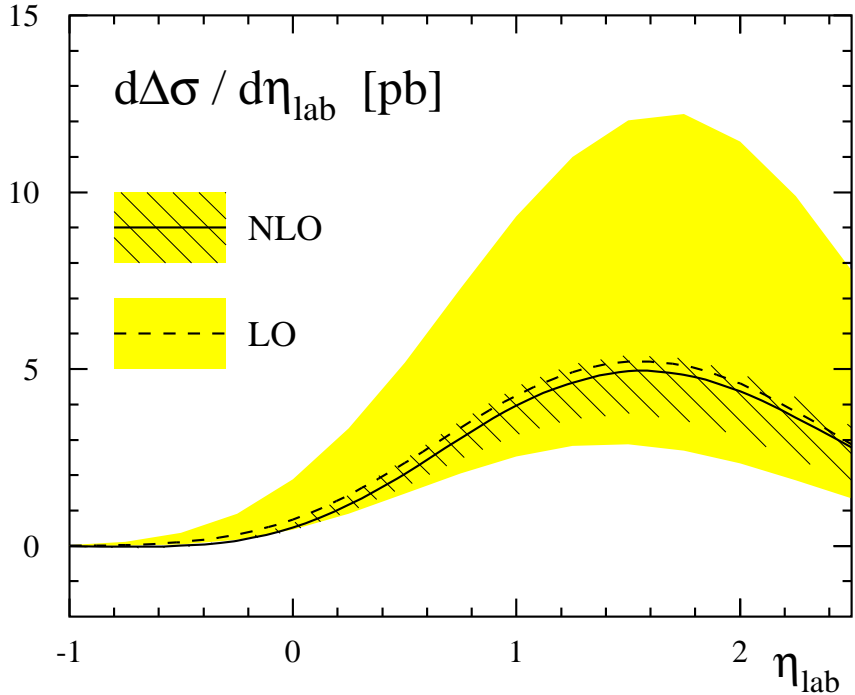,width=0.80\textwidth}
\end{center}
\vspace*{-1.0cm}
\caption{\sf Scale dependence of the polarized cross section
for $\pi^0$ production at LO and NLO in the range 
$p_T/2 \leq \mu_r=\mu_f=\mu_f' \leq 2 p_T$. 
The solid and dashed lines correspond
to the choice where all scales are set to $p_T$.\label{fig:fig3}}
\end{figure}
An important reason why inclusion of the NLO corrections is 
generally vital 
is that they are expected to considerably reduce the dependence of the cross
sections on the unphysical factorization and renormalization scales. 
This improvement in scale dependence when going from LO to NLO is a 
measure of the impact of the NLO corrections, the residual scale dependence
at NLO perhaps providing a rough estimate of the relevance of even 
higher order QCD corrections. Figure~\ref{fig:fig3} shows the scale 
dependence of the spin-dependent cross section at LO and NLO. 
In each case the shaded bands indicate the uncertainties from varying the
unphysical scales in the range $p_T/2 \leq \mu_r=\mu_f=\mu_f' \leq 2 p_T$. 
The solid lines are for the choice where all scales are set to $p_T$.
One can see that the scale dependence indeed becomes much smaller at NLO and
that the lowest order approximation may capture the main features of 
the process but will not provide a satisfactory quantitative understanding. 

%
%
\begin{figure}[t]
\vspace*{-1.2cm}
\begin{center}
\epsfig{figure=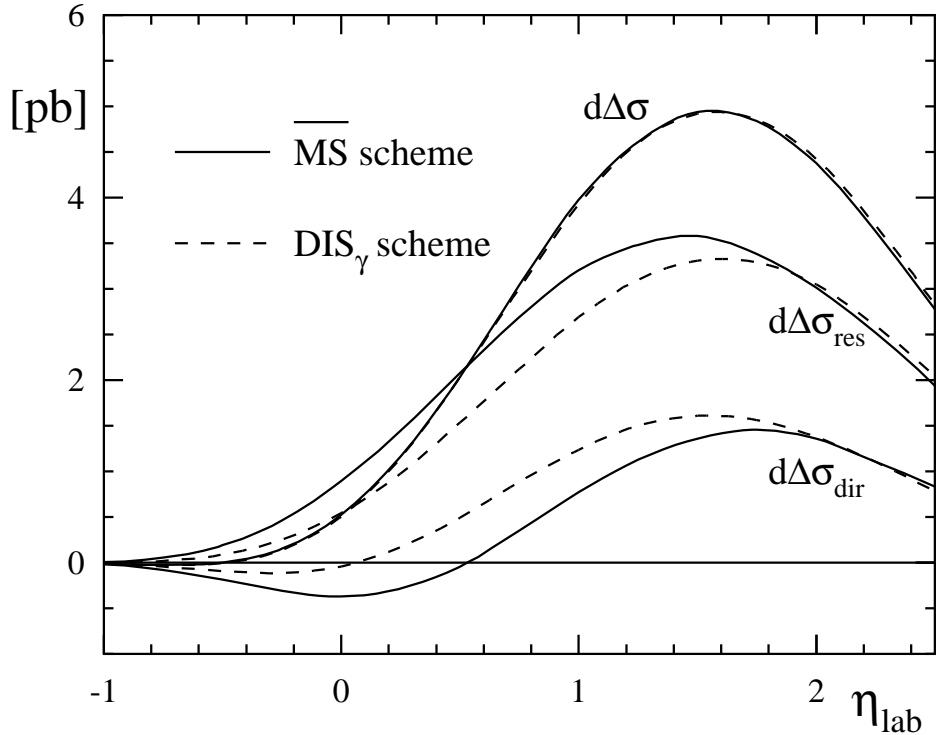,width=0.85\textwidth}
\end{center}
\vspace*{-1cm}
\caption{\sf NLO direct and resolved photon contributions in 
the $\overline{\mathrm{MS}}$ and $\mathrm{DIS}_{\gamma}$ factorization
schemes. Also shown is their sum which has to be independent 
of theoretical conventions. \label{fig:fig4}}
\end{figure}

Figure \ref{fig:fig4} demonstrates, as discussed in Section II, that the 
direct and resolved contributions individually depend on the choice of the 
factorization scheme, but that the physical, i.e., experimentally accessible, 
cross section does not. To make this point, we transform the 
polarized cross sections from the $\overline{\rm{MS}}$ to the 
$\mathrm{DIS}_{\gamma}$ scheme  
as described in Eqs.~(\ref{eq:eq11})-(\ref{eq:eq13}).

%
%
\begin{figure}[t]
\vspace*{-1.2cm}
\begin{center}
\epsfig{figure=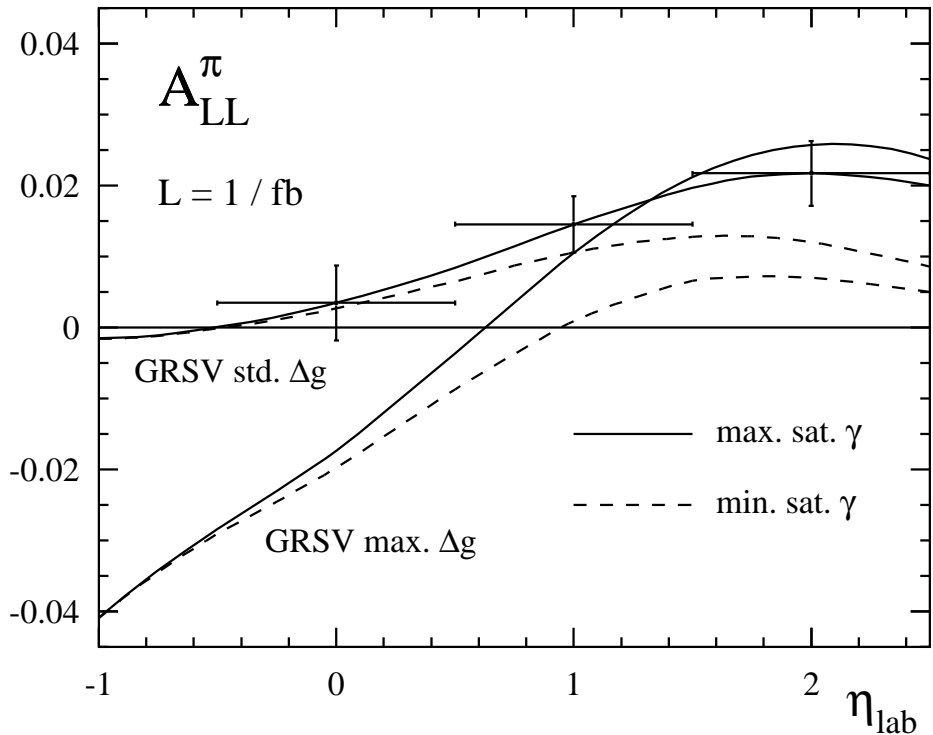,width=0.85\textwidth}
\end{center}
\vspace*{-1cm}
\caption{\sf Spin asymmetry for $\pi^0$ photoproduction in NLO QCD
for the two extreme sets of polarized photon densities and two different
choices of spin-dependent proton distributions. The ``error bars'' 
indicate the statistical accuracy anticipated for eRHIC assuming
an integrated luminosity of $1\,\mathrm{fb}^{-1}$. \label{fig:fig5}}
\end{figure}
 
A first determination of the parton content of circularly polarized photons
could be one of the key physics goals of the spin physics program at a 
future lepton-proton collider. Figure~\ref{fig:fig5} shows the 
experimentally relevant spin asymmetry
\begin{equation}
\label{eq:asydef}
A_{LL}^{\pi}=\frac{d\Delta \sigma}{d\sigma}=\frac{ 
d\sigma_{++} - d\sigma_{+-}}{d\sigma_{++} + d\sigma_{+-}}
\end{equation}
for $\pi^0$ photoproduction at eRHIC. As before we have integrated over
$p_T\geq 4\,\mathrm{GeV}$ and have chosen all scales to be $p_T$.
At positive rapidities the spin asymmetry shows the expected dependence 
on the choice of the polarized photon densities. In order to be able 
to judge whether a future measurement can resolve such a difference we also
give in Figure \ref{fig:fig5} an estimate for the expected statistical
errors:
\begin{equation} \label{error}
\delta A_{LL}^{\pi} \simeq \frac{1}{{\cal{P}}_e {\cal{P}}_p 
\sqrt{{\cal L}\sigma_{\rm bin}}} \; ,
\end{equation}
where ${\cal{P}}_e$ and ${\cal{P}}_p$ are the polarization of 
the lepton and proton beam, respectively, ${\cal L}$ denotes the integrated
luminosity of the collisions, and $\sigma_{\rm bin}$ is the unpolarized
cross section integrated over bins in $\eta_{\mathrm{lab}}$.
We have used ${\cal{P}}_{e,p}=0.7$ and ${\cal L}=1$/fb, 
which are targets for eRHIC. It should be noted that an integrated luminosity
of ${\cal L}=1$/fb is expected to be accumulated after only a few weeks of
running so that the statistical accuracy may eventually be much better than
the one shown in Fig.~\ref{fig:fig5}.

To demonstrate that the sensitivity to $\Delta f^{\gamma}$ is not 
masked by the dependence on the spin-dependent {\em proton} densities, we show 
in Fig.~\ref{fig:fig5} also the results obtained for the GRSV 
``max.\ $\Delta g$'' set \cite{ref:grsv}, which is characterized by
a much larger gluon density $\Delta g$ than the one in the 
GRSV ``standard'' set. It is evident that 
at large positive rapidities the asymmetry is indeed mainly determined
by the polarized photon structure. In the region of $\eta_{\mathrm{lab}}
\le 0$ a measurement of $A_{LL}^{\pi}$ would be also an
excellent source of information on the polarized gluon density in the nucleon.
We note that, in any case, by the time eRHIC is commissioned we expect 
the spin structure of the nucleon to be known with much better accuracy 
than at present. 

%
%
\begin{figure}[t]
\vspace*{-1.2cm}
\begin{center}
\epsfig{figure=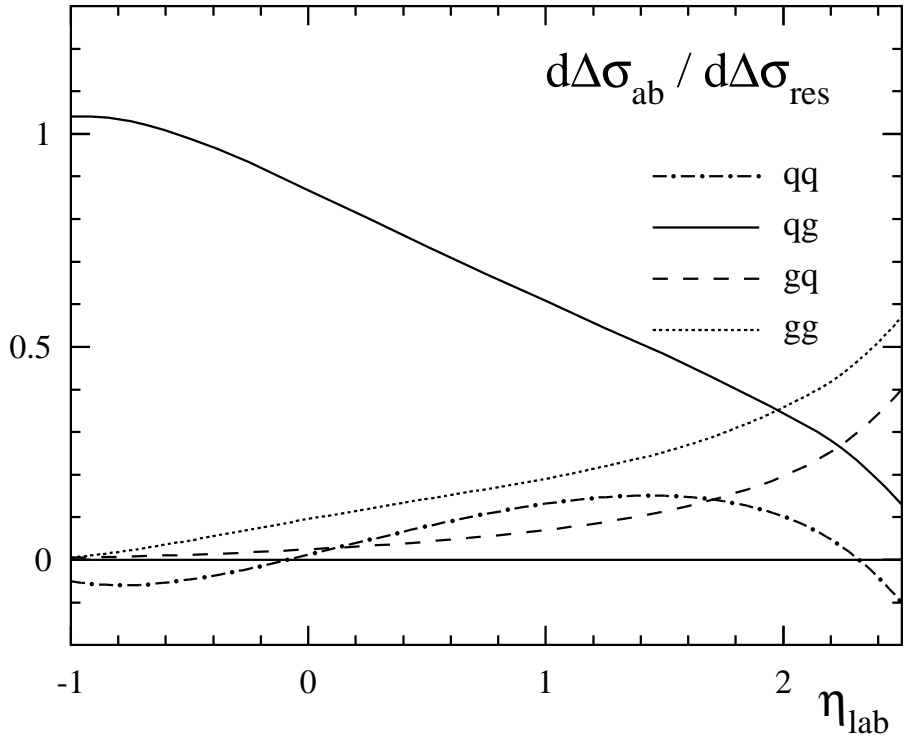,width=0.85\textwidth}
\end{center}
\vspace*{-1cm}
\caption{\sf Relative contributions of different
partonic scatterings $ab\to c X$ to the 
NLO ($\overline{\mathrm{MS}}$) resolved photon cross section 
for $\pi^0$ production at eRHIC, using
the ``standard'' set of GRSV~\cite{ref:grsv} for the proton and
the ``maximal'' scenario for the spin-dependent photon densities. 
\label{fig:fig6}}
\end{figure}

Figure~\ref{fig:fig6} illustrates how different partonic
scatterings $ab\to c X$ contribute to the  NLO ($\overline{\mathrm{MS}}$) 
polarized resolved cross section, versus rapidity. Recall that parton 
``$a$'' emerges from the photon and
parton ``$b$'' from the proton. Qualitatively similar results are 
obtained for the ``minimal'' photon scenario and hence not shown here.

\section{Conclusions}
%
In this paper we have presented the first complete NLO calculation for the
spin-dependent photoproduction of inclusive high-$p_T$ hadrons, consistently including
the direct and resolved photon contributions at NLO.  Our results are 
relevant for future polarized lepton-proton collider projects like eRHIC,
to which we have tailored our numerical results. 
The spin asymmetry for this process is a promising tool for a first 
determination of the spin structure of circularly polarized photons.
We found that the QCD corrections to the polarized cross section
are under good control. The polarized cross section shows a significant 
reduction of scale dependence when going from LO to NLO. 

We finally note that a single process alone, like pion photoproduction,
will allow to establish the very existence of a polarized resolved-photon 
contribution and may perhaps point towards one of the two scenarios discussed
above. It will, however, not be sufficient to fully disentangle 
the various quark flavor distributions 
and the gluon density of the polarized photon. 
This will require, as for the nucleon (spin) structure, a multitude of 
processes, combined in a global QCD analysis. Of course, at eRHIC other 
photoproduction processes, like jets, di-jets, or heavy flavors, 
will add further valuable information. Measurements at a future polarized 
linear $e^+e^-$ collider would be also extremely helpful in this respect.

\section*{Acknowledgments}
%
We are grateful to A.\ Deshpande for many valuable discussions about the 
eRHIC project. We thank A.\ Sch\"{a}fer for discussions. 
B.J.\ and M.S.\ thank the RIKEN-BNL Research Center 
for hospitality and support during the final stages of this work, 
and they acknowledge support from the Program Development 
Fund PD02-98 of Brookhaven Science Associates.
B.J.\ was supported by the European Commission IHP program under contract number
HPRN-CT-2000-00130. 
W.V.\ is grateful to RIKEN, Brookhaven National Laboratory and the U.S.\
Department of Energy (contract number DE-AC02-98CH10886) for
providing the facilities essential for the completion of this work.
This work was supported in part by the ``Bundesministerium f\"{u}r
Bildung und Forschung (BMBF)'' and the ``Deutsche 
Forschungsgemeinschaft (DFG)''.

%

%

\newpage
\begin{thebibliography}{99}
%
\bibitem{ref:rhic} See, for example: G.\ Bunce, N.\ Saito, J.\ Soffer, and 
W.\ Vogelsang, Annu.\ Rev.\ Nucl.\ Part.\ Sci.\ {\bf 50}, 525 (2000).
%
\bibitem{ref:owens} See discussions in: J.F.\ Owens, Phys. Rev. {\bf D21}, 54
(1980); \\
M.\ Drees and R.M.\ Godbole, Phys. Rev. Lett. {\bf 61}, 682 (1988);
Phys. Rev. {\bf D39}, 169 (1989).
%
\bibitem{ref:klasen} A recent review can be found, e.g., in: M.\ Klasen, 
Rev. Mod. Phys. {\bf 74}, 1221 (2002).
%
\bibitem{ref:eic} See {\tt http://www.bnl.gov/eic} for information 
concerning the eRHIC/EIC project, including the ``Whitepaper'' BNL-68933.
%
\bibitem{ref:lostudies}  M.\ Stratmann and W.\ Vogelsang, 
Z. Phys. {\bf C74}, 641 (1997);\\
J.M.\ Butterworth, N.\ Goodman, M.\ Stratmann, and W.\ Vogelsang, in
proceedings of the workshop on ``Physics with Polarized Protons at HERA'',
Hamburg, Germany, 1997, p.\ 120 [{\tt hep-ph/9711250}]. 
%
\bibitem{ref:nlopion} B.\ J\"{a}ger, A.\ Sch\"{a}fer, M.\ Stratmann,
and W.\ Vogelsang, Phys. Rev. {\bf D67}, 054005 (2003).
%
\bibitem{ref:danielpion} D.\ de Florian, Phys. Rev. {\bf D67}, 054004 (2003).
%
\bibitem{ref:nlopoldir}  D.\ de Florian and W.\ Vogelsang, 
Phys. Rev. {\bf D57}, 4376 (1998). 
%
\bibitem{ref:daniel} D.\ de Florian and S.\ Frixione, Phys. Lett. {\bf B457}, 
236 (1999).
%
\bibitem{ref:aversa} F.\ Aversa, P.\ Chiappetta, M.\ Greco, and 
J.-Ph.\ Guillet, Nucl. Phys. {\bf B327}, 105 (1989).
%
\bibitem{ref:aurenche} P.\ Aurenche, R.\ Baier, A.\ Douiri, M.\ Fontannaz,
and D.\ Schiff, Nucl. Phys. {\bf B286}, 553 (1987).
%
\bibitem{ref:nlounpdir} L.E.\ Gordon, Phys. Rev. {\bf D50}, 6753 (1994).
%
\bibitem{ref:grvdisgamma} M.\ Gl\"{u}ck, E.\ Reya, and A.\ Vogt, 
Phys. Rev. {\bf D45}, 3986 (1992).
%
\bibitem{ref:polphotnlo}  M.\ Stratmann and W.\ Vogelsang, Phys. Lett. 
{\bf B386}, 370 (1996).
%
\bibitem{ref:cteq5} CTEQ Collaboration, H.-L. Lai {\it et al.}, 
Eur. Phys. J. {\bf C12}, 375 (2000).
%
\bibitem{ref:grsv}  M.\ Gl\"{u}ck, E.\ Reya, M.\ Stratmann, and
W.\ Vogelsang, Phys. Rev. {\bf D63}, 094005 (2001). 
%
\bibitem{ref:kkp} B.A.\ Kniehl, G.\ Kramer, and B.\ P\"{o}tter, Nucl. Phys. 
{\bf B582}, 514 (2000).
%
\bibitem{ref:kretzer} S.\ Kretzer, Phys. Rev. {\bf D62}, 054001 (2000).
%
\bibitem{ref:phenixpion} PHENIX Collaboration, S.S.\ Adler 
{\em et al.}, {\tt hep-ex/0304038}.
%
\bibitem{ref:lomodels} M.\ Gl\"{u}ck and W.\ Vogelsang, Z. Phys. 
{\bf C55}, 353 (1992); {\em ibid.} {\bf C57}, 309 (1993);\\
M.\ Gl\"{u}ck, M.\ Stratmann, and W.\ Vogelsang, Phys. Lett. {\bf B337}, 
373 (1994).
%
\bibitem{ref:grvphot} M.\ Gl\"{u}ck, E.\ Reya, and A.\ Vogt, 
Phys. Rev. {\bf D46}, 1973 (1992).
%
\bibitem{ref:sieg} M.\ Gl\"{u}ck, E.\ Reya, and C.\ Sieg, Phys. Lett. 
{\bf B503}, 285 (2001).
%
%
\end{thebibliography}
\end{document}